\documentclass[twocolumn]{aastex63}
\usepackage{apjfonts, natbib, lineno}
\usepackage{longtable, booktabs, array, amsmath, subfigure}

\newcommand{\lum}{erg~s\ensuremath{^{-1}}}

\newcommand{\msun}{\ensuremath{M_{\odot}}}

\newcommand{\mbh}{\ensuremath{M_\mathrm{BH}}}
\newcommand{\wise}{\emph{WISE}}
\newcommand{\neowise}{\emph{NEOWISE}}

\newcommand{\tdust}{\ensuremath{T_\mathrm{dust}}}

\newcommand{\ldust}{\ensuremath{L_\mathrm{dust}}}

\newcommand{\lbb}{\ensuremath{L_\mathrm{bb}}}
\newcommand{\tbb}{\ensuremath{T_\mathrm{BB}}}

\shorttitle{IR echo of TDE}
\shortauthors{Jiang et al.}

\begin{document}
\title{Infrared Echoes of Optical Tidal Disruption Events: 
$\sim$1\% Dust Covering Factor or Less at sub-parsec Scale}

\correspondingauthor{Ning Jiang}
\email{jnac@ustc.edu.cn}

\author[0000-0002-7152-3621]{Ning Jiang}
\affiliation{CAS Key laboratory for Research in Galaxies and Cosmology,
Department of Astronomy, University of Science and Technology of China,
Hefei, 230026, China; jnac@ustc.edu.cn}
\affiliation{School of Astronomy and Space Sciences,
University of Science and Technology of China, Hefei, 230026, China; twang@ustc.edu.cn}

\author[0000-0002-1517-6792]{Tinggui Wang}
\affiliation{CAS Key laboratory for Research in Galaxies and Cosmology,
Department of Astronomy, University of Science and Technology of China,
Hefei, 230026, China; jnac@ustc.edu.cn}
\affiliation{School of Astronomy and Space Sciences,
University of Science and Technology of China, Hefei, 230026, China; twang@ustc.edu.cn}

\author{Xueyang~Hu}
\affiliation{CAS Key laboratory for Research in Galaxies and Cosmology,
Department of Astronomy, University of Science and Technology of China,
Hefei, 230026, China; jnac@ustc.edu.cn}
\affiliation{School of Astronomy and Space Sciences,
University of Science and Technology of China, Hefei, 230026, China; twang@ustc.edu.cn}

\author[0000-0002-7223-5840]{Luming~Sun}
\affiliation{Department of Physics, Anhui Normal University, Wuhu, Anhui, 241000, People's Republic of China}

\author[0000-0002-4757-8622]{Liming Dou}
\affiliation{Department of Astronomy, Guangzhou University, Guangzhou 510006, China}

\author[0000-0002-6986-5593]{Lin~Xiao} 
\affiliation{CAS Key laboratory for Research in Galaxies and Cosmology,
Department of Astronomy, University of Science and Technology of China,
Hefei, 230026, China; jnac@ustc.edu.cn}
\affiliation{School of Astronomy and Space Sciences,
University of Science and Technology of China, Hefei, 230026, China; twang@ustc.edu.cn}

\begin{abstract}

The past decade has experienced an explosive increase of optically-discovered
tidal disruption events (TDEs) with the advent of modern time-domain surveys.
However, we still lack a comprehensive observational view of 
their infrared (IR) echoes in spite of individual detections. 
To this end, we have conducted a statistical study of IR variability of the 23 
optical TDEs discovered between 2009 and 2018 utilizing the full public
dataset of \emph{Wide-field Infrared Survey Explorer}.
The detection of variability is performed on the difference images, 
yielding out 11 objects with significant (>$3\sigma$) variability 
in at least one band while dust emission can be only fitted in 8 objects.
Their peak dust luminosity is around $10^{41}-10^{42}$~\lum,
corresponding to a dust covering factor $f_c\sim0.01$ at scale of sub-parsec.
The only exception is the disputed source ASASSN-15lh, which shows an 
ultra-high dust luminosity ($\sim10^{43.5}$~\lum) and make its nature even elusive.
Other non-detected objects show even lower $f_c$, which could be one more 
order of magnitude lower.
The derived $f_c$ is generally much smaller than those of dusty tori in 
active galactic nuclei (AGNs), suggesting either a dearth of dust or 
a geometrically thin and flat disk in the vicinity of SMBHs.
Our results also indicate that the optical TDE sample (post-starburst
galaxies overrepresented) is seriously biased to events with little dust 
at sub-pc scale while TDEs in dusty star-forming systems 
could be more efficiently unveiled by IR echoes.

\end{abstract}
	

\section{Introduction} 
\label{sec:intro}

Supermassive black holes (SMBHs) are ubiquitous in the centers 
of galaxies with massive bulges. Moreover, the tightness of the BH-bulge mass 
relationship hints a symbiotic connection between the formation and growth 
of BHs and galaxy spheroids (\citealt{KH2013}). 
The SMBHs accumulate their tremendous mass ($10^6-10^{10}$~\msun)
by accreting gas through the phase of active galactic nucleus (AGNs)
while they are mostly quiescent in the local universe.
A fundamental question has been raised why some SMBHs are 
active but the majority of remaining are not (e.g., \citealt{Alexander2012}).
Observational constrains on the AGN triggering mechanisms remain elusive 
with only sparse evidence at galactic scales (\citealt{Storchi-Bergmann2019}).
As the accreting material, the interstellar medium (ISM) at different scales
might give important clues to the underlying mechanism.
It is found that Seyfert AGNs generally reside in host galaxies 
with a younger stellar population than quiescent galaxies, confirming that 
an abundant fuel supply is available in kiloparsec (kpc) 
scale (e.g., \citealt{Kauffmann2003}). 
Further efforts are devoted to explore their differences at smaller scale, 
such as finding a factor of four difference in gas mass between 
Seyfert and quiescent galaxies within radius of 100 pc (\citealt{Hicks2013}).
Nevertheless, a clear picture of AGN triggering is unachievable without
going deep into the proximity of SMBHs, that is down to $pc$-scale under the 
gravitational influence of SMBHs.

The structure of AGNs is cognized under the scheme of 
unified model (\citealt{Antonucci1993}), 
in which the equatorial optically thick torus at $pc$-scale 
lays the foundation of unification 
and has bridged the scale of accretion disk and their host galaxy.
Besides, the dust along the polar direction may also exist in some AGNs 
suggested by interferometric observations (\citealt{Honig2012}; 
see also \citealt{Lyu2018}), leading to the dusty wind model 
(\citealt{Honig2013}; \citealt{Honig2017}).
In contrast, the study of $pc$-scale environment of normal galaxies is much more 
challenging without the illumination from central engines.
The far-IR imaging of our Milky Way center
has uncovered a circumnuclear ring centered on Sgr~A$^\star$ with 
thickness and radius of 0.34~pc and 1.4~pc, respectively 
(\citealt{Lau2013}; see also \citealt{Latvakoski1999} for an agreed result).
However, similar map is impossible for more distant galaxies due to poor resolution.
The groundbreaking instrument GRAVITY mounted on the very large telescope
has achieved mili-arcsec resolution in $K$-band, but it only applies to 
$K$-band luminous sources and thus barely nearby AGNs have been successfully 
observed (\citealt{Gravity2020a,Gravity2020b}).

Nevertheless, the gas and dust in the vicinity of inactive SMBHs have another 
possibility to be lighted up temporarily by tidal disruption events (TDEs),
which happens when a star occasionally passes within the tidal radius of SMBH.
Part of the disrupted stellar debris will be accreted by the BH
and produce a flash of electromagnetic radiation peaked in UV or soft X-ray band, 
with a characteristic  $t^{-5/3}$ decline on timescale of months to years 
(\citealt{Rees1988,Phinney1989}).
If the local environment of a SMBH is dusty, the UV/optical photons 
from TDE will be unavoidably absorbed and re-processed into 
the infrared (IR) band, giving rise to a so-called IR echo.
\citet{Lu2016} has performed a 1-D radiative transfer model and 
proved that the dust emission peaks at mid-IR (MIR) with
typical luminosity between $10^{42}$ and $10^{43}$~\lum\
depending on the dust covering factor.
Immediately after the prediction, \citet{Jiang2016} achieved the detection 
of IR echo of dust at scale of $\sim0.1pc$ in ASASSN-14li. 
At almost the same time,  \citet{vV2016} have claimed another two echoes 
(PTF-09ge and PTF-09axc) and derived a dust covering factor of $\sim1\%$ 
for both events.
Therefore, IR echoes of TDEs offer us a new and powerful means to probe 
the dust around the BH down to sub-pc scale.

The technique of IR echo is subject to the occurrence of TDEs.
The TDE event rate is yet known to be as low as $10^{-4}-10^{-5}/galaxy/year$
(\citealt{Wang2004,Stone2016}) and thus makes the discovered events quite rare
until recently. The number of TDEs has been growing rapidly 
in the past decade thanks to the booming wide-field optical surveys, 
such as PanSTARRS, PTF, and ASAS-SN (\citealt{Komossa2015,vV2020}).
Particularly, the ZTF survey (\citealt{Graham2019}) since 2018 has made 
TDE enter into a new era of population studies (\citealt{vV2021}).  
As of the end of 2019, approximately 30 TDEs have been discovered 
at optical bands (see Table 1 of \citealt{vV2020}). 
On the other side, the detection of IR echoes all have 
used the archival data of 
\emph{Wide-field Infrared Survey Explorer} (\wise; \citealt{Wright2010}) and
Near-Earth Object \wise\ Reactivation mission (\neowise-R; \citealt{Mainzer2014}).
Actually, the dataset has provided multi-epoch MIR date with 
time coverage matching with almost all optical TDEs.
There is however still no statistical study to date.
Giving the increased number of optical TDEs in the past few years and 
available IR data, it is the perfect time to perform a comprehensive study.
We assume a cosmology with $H_{0} =70$ km~s$^{-1}$~Mpc$^{-1}$, $\Omega_{m} = 
0.3$, and $\Omega_{\Lambda} = 0.7$.

\begin{figure*}
\centering
\includegraphics[width=15cm]{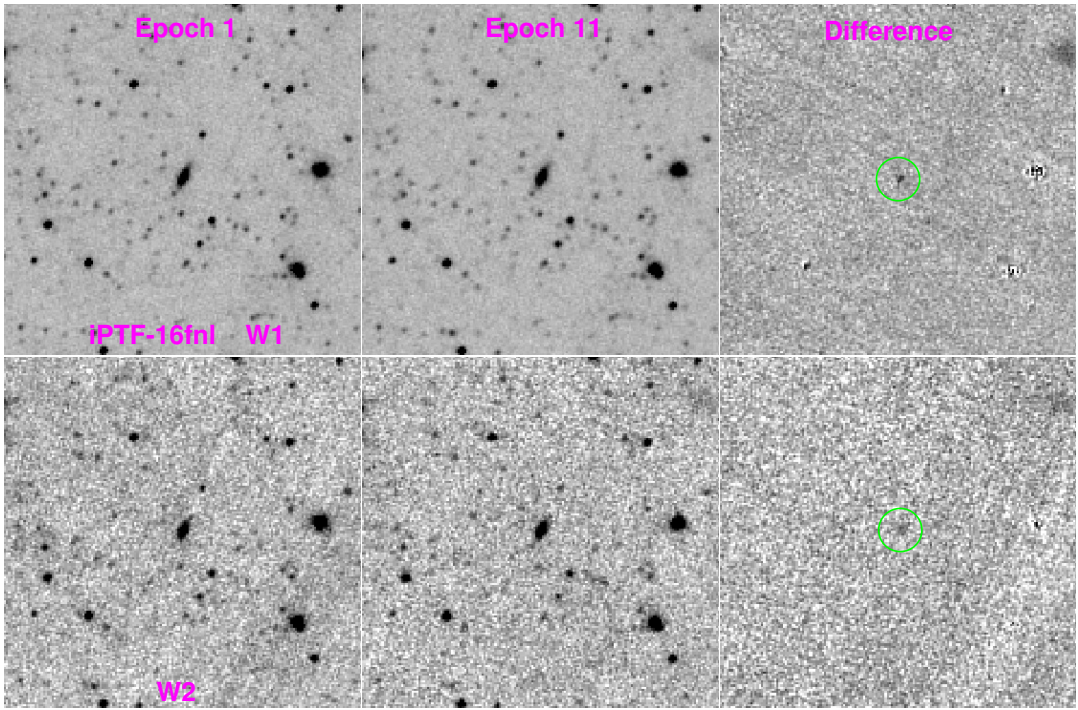}
\caption{
We show iPTF-16fnl as an example of image subtraction.
The IR variability of this TDE are invisible from the original \wise\ light curves
while they are robustly detected with PSF photometrry 
in the difference images (epoch 1 as reference).
}
\label{iptf16fnl}
\end{figure*}

\begin{deluxetable*}{clcclcccccccc}
\tabletypesize{\scriptsize}
\setlength{\tabcolsep}{0.06in}
\tablecaption{\it Sample of Optical TDEs}
\label{sample}
\tablewidth{0pt}
\tablehead{
	\colhead{ID} & \colhead{Name} & \colhead{IAU Name} & \colhead{R.A.} & \colhead{DEC.} & \colhead{$z$} & \colhead{$\rm MJD_{peak}$} & \colhead{log\lbb} & \colhead{log<\tbb>} & \colhead{log~$t_0$} & \colhead{p} & \colhead{log\mbh}  & \colhead{log$M_{\star}$} \\ 
&  &  &  &  &  &  & \colhead{erg/s} & \colhead{K} & \colhead{Day}  &  & \colhead{\msun} & \colhead{\msun}  \\
\colhead{(1)} & \colhead{(2)}  & \colhead{(3)}  & \colhead{(4)} & \colhead{(5)} & 
\colhead{(6)} & \colhead{(7)}  & \colhead{(8)} & \colhead{(9)} & \colhead{(10)}  & \colhead{(11)} & \colhead{(12)} & \colhead{(13)}
} 
\startdata
1  & PTF-09axc    &  ...      & 14:53:13.08 & +22:14:32.3 & 0.1146 & 55016 & $43.46^{+0.03}_{-0.02}$ & $4.08^{+0.00}_{-0.00}$ & $2.21^{+0.14}_{-0.17}$ & $-1.5^{+0.3}_{-0.3}$ & $5.68^{+0.48}_{-0.49}$ & $10.12^{+0.11}_{-0.17}$ \\ 
2  & PTF-09djl    &  ...      & 16:33:55.97 & +30:14:16.7 & 0.1840 & 55048 & $44.42^{+0.04}_{-0.04}$ & $4.41^{+0.00}_{-0.00}$ & $1.67^{+0.11}_{-0.14}$ & $-1.7^{+0.2}_{-0.3}$ & $5.82^{+0.56}_{-0.58}$ & $ 9.95^{+0.15}_{-0.12}$ \\ 
3  & PTF-09ge     &  ...      & 14:57:03.18 & +49:36:41.1 & 0.0640 & 54992 & $44.04^{+0.01}_{-0.01}$ & $4.31^{+0.03}_{-0.03}$ & $1.85^{+0.08}_{-0.10}$ & $-1.7^{+0.2}_{-0.2}$ & $6.31^{+0.39}_{-0.39}$ & $10.11^{+0.13}_{-0.12}$ \\ 
4  & PS1-10jh     &  ...      & 16:09:28.28 & +53:40:24.1 & 0.1696 & 55393 & $44.47^{+0.07}_{-0.07}$ & $4.49^{+0.03}_{-0.03}$ & $1.44^{+0.41}_{-0.22}$ & $-1.5^{+0.2}_{-0.6}$ & $5.85^{+0.44}_{-0.44}$ & $ 9.61^{+0.10}_{-0.13}$ \\ 
5  & PS1-11af     &  ...      & 09:57:26.82 & +03:14:00.9 & 0.4046 & 55578 & $44.16^{+0.03}_{-0.03}$ & $4.31^{+0.01}_{-0.01}$ & $1.73^{+0.16}_{-0.17}$ & $-2.1^{+0.5}_{-0.5}$ & $6.57^{+0.06}_{-0.05}$ & $10.21^{+0.21}_{-0.19}$ \\ 
6  & ASASSN-14ae  &  ...      & 11:08:40.12 & +34:05:52.2 & 0.0436 & $56684^*$ & $43.87^{+0.02}_{-0.01}$ & $4.27^{+0.01}_{-0.01}$ & $1.55^{+0.05}_{-0.07}$ & $-2.4^{+0.2}_{-0.1}$ & $5.42^{+0.46}_{-0.46}$ & $10.02^{+0.09}_{-0.17}$ \\ 
7  & ASASSN-14li  &  ...      & 12:48:15.23 & +17:46:26.4 & 0.0206 & $56989^*$ & $43.66^{+0.02}_{-0.02}$ & $4.51^{+0.01}_{-0.01}$ & $1.69^{+0.09}_{-0.09}$ & $-1.4^{+0.1}_{-0.1}$ & $6.23^{+0.39}_{-0.40}$ & $ 9.69^{+0.05}_{-0.10}$ \\ 
8  & ASASSN-15oi  &  ...      & 20:39:09.18 & -30:45:20.1 & 0.0484 & $57178^*$ & $44.47^{+0.04}_{-0.04}$ & $4.52^{+0.02}_{-0.02}$ & $1.69^{+0.08}_{-0.08}$ & $-2.4^{+0.2}_{-0.2}$ & $6.60^{+0.10}_{-0.12}$ & $10.02^{+0.04}_{-0.04}$ \\ 
9  & ASASSN-15lh  &  ...      & 22:02:15.39 & -61:39:34.6 & 0.2326 & 57247 & $45.34^{+0.04}_{-0.04}$ & $4.20^{+0.03}_{-0.03}$ & ... & ... & $8.72^{+0.40}_{-0.36}$ & $10.95^{+0.15}_{-0.11}$ \\ 
10 & iPTF-15af    &  ...      & 08:48:28.13 & +22:03:33.4 & 0.0790 & 57061 & $44.22^{+0.01}_{-0.01}$ & $4.70^{+0.04}_{-0.03}$ & $1.91^{+0.22}_{-0.22}$ & $-1.5^{+0.3}_{-0.3}$ & $6.88^{+0.38}_{-0.38}$ & $10.12^{+0.11}_{-0.17}$ \\ 
11 & iPTF-16axa   &  ...      & 17:03:34.34 & +30:35:36.6 & 0.1080 & $57523^*$ & $43.82^{+0.02}_{-0.01}$ & $4.37^{+0.01}_{-0.01}$ & $1.54^{+0.13}_{-0.16}$ & $-1.6^{+0.3}_{-0.2}$ & $6.34^{+0.42}_{-0.42}$ & $10.18^{+0.10}_{-0.14}$ \\ 
12 & iPTF-16fnl   &  ...      & 00:29:57.01 & +32:53:37.2 & 0.0163 & 57626 & $43.18^{+0.03}_{-0.02}$ & $4.08^{+0.00}_{-0.00}$ & $1.35^{+0.08}_{-0.09}$ & $-2.1^{+0.2}_{-0.2}$ & $5.50^{+0.42}_{-0.42}$ & $ 9.35^{+0.12}_{-0.15}$ \\ 
13 & OGLE16aaa    &  ...      & 01:07:20.81 & -64:16:21.4 & 0.1655 & 57414 & $44.22^{+0.01}_{-0.01}$ & $4.23^{+0.00}_{-0.00}$ & $2.19^{+0.09}_{-0.09}$ & $-2.2^{+0.3}_{-0.4}$ & $6.48^{+0.15}_{-0.13}$ & $10.47^{+0.09}_{-0.11}$ \\ 
14 & PS17dhz      & AT2017eqx & 22:26:48.37 & +17:08:52.4 & 0.1089 & 57910 & $43.82^{+0.03}_{-0.05}$ & $4.31^{+0.01}_{-0.01}$ & $1.47^{+0.10}_{-0.13}$ & $-2.0^{+0.3}_{-0.3}$ & $5.81^{+0.22}_{-0.23}$ & $ 9.44^{+0.11}_{-0.13}$ \\ 
15 & PS18kh       & AT2018zr  & 07:56:54.54 & +34:15:43.6 & 0.0710 & 58180 & $43.78^{+0.02}_{-0.02}$ & $4.14^{+0.01}_{-0.01}$ & $1.23^{+0.15}_{-0.12}$ & $-0.8^{+0.0}_{-0.1}$ & $6.43^{+0.16}_{-0.23}$ & $10.03^{+0.09}_{-0.18}$ \\ 
16 & ASASSN-18pg  & AT2018dyb & 16:10:58.77 & -60:55:23.2 & 0.0180 & 58346 & $44.16^{+0.01}_{-0.01}$ & $4.37^{+0.01}_{-0.01}$ & $1.49^{+0.05}_{-0.05}$ & $-1.9^{+0.1}_{-0.1}$ & $6.40^{+0.18}_{-0.21}$ & $10.00^{+0.11}_{-0.15}$ \\ 
17 & ASASSN-18ul  & AT2018fyk & 22:50:16.13 & -44:51:52.4 & 0.0590 & $58317^*$ & $44.48^{+0.04}_{-0.03}$ & $4.56^{+0.02}_{-0.02}$ & $2.14^{+0.06}_{-0.06}$ & $-1.9^{+0.1}_{-0.1}$ & $7.00^{+0.16}_{-0.18}$ & $10.57^{+0.12}_{-0.15}$ \\ 
18 & ASASSN-18zj  & AT2018hyz & 10:06:50.87 & +01:41:34.1 & 0.0457 & $58427^*$ & $44.11^{+0.01}_{-0.01}$ & $4.25^{+0.01}_{-0.01}$ & $1.29^{+0.07}_{-0.06}$ & $-1.1^{+0.1}_{-0.1}$ & $6.14^{+0.20}_{-0.32}$ & $ 9.75^{+0.12}_{-0.26}$ \\ 
19 & ZTF19aabbnzo & AT2018lna & 07:03:18.65 & +23:01:44.7 & 0.0910 & 58508 & $44.56^{+0.06}_{-0.06}$ & $4.60^{+0.03}_{-0.02}$ & $1.79^{+0.24}_{-0.21}$ & $-2.1^{+0.7}_{-0.7}$ & $5.84^{+0.22}_{-0.21}$ & $ 9.47^{+0.12}_{-0.09}$ \\ 
20 & ZTF18aahqkbt & AT2018bsi & 08:15:26.62 & +45:35:32.0 & 0.0510 & $58217^*$ & $43.87^{+0.01}_{-0.01}$ & $4.37^{+0.03}_{-0.03}$ & $1.92^{+0.25}_{-0.21}$ & $-1.9^{+0.4}_{-0.6}$ & $7.04^{+0.10}_{-0.11}$ & $10.61^{+0.05}_{-0.06}$ \\ 
21 & ZTF18abxftqm & AT2018hco & 01:07:33.64 & +23:28:34.3 & 0.0880 & 58401 & $44.25^{+0.04}_{-0.04}$ & $4.39^{+0.01}_{-0.01}$ & $1.73^{+0.16}_{-0.15}$ & $-1.2^{+0.2}_{-0.2}$ & $6.29^{+0.17}_{-0.24}$ & $ 9.90^{+0.09}_{-0.18}$ \\ 
22 & ZTF18acaqdaa & AT2018iih & 17:28:03.93 & +30:41:31.4 & 0.2120 & 58442 & $44.62^{+0.04}_{-0.03}$ & $4.23^{+0.01}_{-0.01}$ & $1.61^{+0.11}_{-0.07}$ & $-0.9^{+0.1}_{-0.1}$ & $7.20^{+0.13}_{-0.18}$ & $10.76^{+0.09}_{-0.15}$ \\ 
23 & ZTF18actaqdw & AT2018lni & 04:09:37.65 & +73:53:41.7 & 0.1380 & 58460 & $44.21^{+0.29}_{-0.17}$ & $4.44^{+0.09}_{-0.07}$ & $2.46^{+0.37}_{-0.45}$ & $-1.4^{+0.8}_{-1.6}$ & $6.34^{+0.18}_{-0.21}$ & $ 9.94^{+0.10}_{-0.15}$    
\enddata
\tablecomments{
Column~(1): Object ID in this paper.
Column~(2): Discovery Name of the TDE.
Column~(3): IAU Name of the TDE.
Column~(4)-(5): RA and DEC of the TDE.
Column~(6): redshift of the TDE host galaxy.
Column~(7): MJD of the optical peak. The asterisks indicate the time of 
fist detection since these TDEs are only detected post-peak.
Column~(8): the UV-optical bolometric luminosity (\lbb)  the optical peak which is drawn from \citet{vV2020}.
Column~(9): mean blackbody temperature (\tbb) measured during the first 100 days post peak.
Column~(10): linear \tbb change during the first year of observations.
Column~(10)-(11): $p$ and $t_0$ are the free parameters of a power-law decay 
(\lbb$\propto(t/t_{0})^p$).
Column~(12): black hole mass (\mbh) derived from 
\mbh-$\sigma_{\star}$ relation for target  1,2,3,4,6,7,10,11,12 (\citealt{Wevers2017}) and 9 (\citealt{Kruhler2018});
from model fitting for target 5,8,13 (\citealt{Mockler2019}); from \mbh-$M_{\star}$ for other objects (\citealt{Reines2015}).
Column~(13): host stellar mass.
The data from Column (7) to (11) are all drawn from \citet{vV2020} except for ASASSN-15lh.
}
\end{deluxetable*}

\section{Sample and Data}
\label{TDEsample}

\subsection{TDE Sample}

The TDE candidates studied in this work are primarily collected from 
\citet{vV2020}, which has reviewed all TDEs discovered in optical band up to 2019.
We have only selected optical TDEs since they usually
possess well-sampled multi-wavelength light curves with wide time span
which is important for us to get the knowledge of the basic
properties of these events, such as the peak time and luminosity.
Furthermore, we only investigate events found between 2009 and 2018 to ensure
available MIR data within one year after the TDE, since the \wise\ project 
starts from early 2010 and its public data goes on to the end of 2019.  
The cut results in 22 sources.
In addition, we have also included another controversial TDE candidate 
ASASSN-15lh (\citealt{Leloudas2016}; \citealt{Kruhler2018}) which was first
reported as the most ever luminous supernova (\citealt{Dong2016}).
We take it into consideration in hope of gaining
some new clues from the IR variability.
Thus, our final sample has 23 objects in total 
(see their information in Table~\ref{sample}).

\subsection{MIR Data}

The \wise\ has conducted a full-sky imaging survey in four broad MIR
bandpass filters centered at 3.4, 4.6, 12 and 22~$\mu$m (labeled W1-W4)
from 2010 February to August (\citealt{Wright2010}).
The solid hydrogen cryogen used to cool the W3 and W4 instrumentation
was depleted later and it was placed in hibernation in 2011 February.
\wise\ was reactivated and renamed \neowise-R since 2013 October, 
using only W1 and W2, to hunt for asteroids that could pose as impact hazard 
to the Earth (\citealt{Mainzer2014}).
The \wise\ scans a specific sky area every half year and average 12 times 
of single exposures have been taken within each epoch (typically one day). 
As of now, the \wise\ and its successor \neowise\ surveys 
have provided us a public dataset from 2010 February to 2019 December, 
which contains 14-15 epochs of observations for each TDE.
Therefore, the observing schedule of \wise\ is in excellent overlap with 
the discovery period of the optical TDEs in our sample (2009-2018).

Our previous works have shown that the IR echoes of TDEs are detectable 
on time scales of months to years while the variability within each epoch 
is negligible (\citealt{Jiang2016,Jiang2017,Jiang2019,Dou2016,Dou2017}), 
so the original single-exposure photometry have been simply binned in those works. 
However, it is not accurate enough to detect weak variability 
or put clear upper limit of the non-detection sources. 
In order to acquire more accurate measurements, 
particularly for TDEs with weak echoes,
We choose to perform photometry on the time-resolved \wise/\neowise\ Coadds.
The coadds have stacked the individual exposures within 
typically $\sim1$ day intervals 
to produce one coadd per band per epoch$--$that is, one coadd every six months 
at a given position on the sky (\citealt{Meisner2018})\footnote{
Website link:
https://portal.nersc.gov/project/cosmo/temp/ameisner/neo6}. 
In addition, the associated noise and mask images have been also generated 
during the process. 
Therefore, it provides us a convenient dataset which is very
suitable for study the long-timescale MIR variability.

\section{Analysis and Results}

\subsection{Variability Detection by Image Subtraction}

We try to detect variability using the standard image subtraction procedure
{\tt HOTPANTS} (\citealt{Becker2015})~\footnote{
https://github.com/acbecker/hotpants}.
The images at latest epoch are taken as the references to be subtracted 
for TDEs discovered before 2013, otherwise the first epoch images are adopted.
Then we begin to perform PSF photometry on the difference images using the {\tt IDL}
routine {\tt FASTPHOT} (\citealt{Bethermin2010}). 
The PSF models specifically constructed for coadd images 
(\citealt{Meisner2019})~\footnote{https://github.com/legacysurvey/unwise\_psf}
have been used as the input PSF images during the measurement. 

Since the single-epoch reference image we used above can be slightly offset 
from the real quiescent level from the host galaxy emission, 
we begin to estimate the offset by averaging fluxes of epochs at least 180 days 
before the optical peak for TDEs after 2013, 
or at least 1500 days after optical peak for TDEs before 2013.
Then we corrected the offset and added its error to the fluxes of difference images.
Finally, we have obtained the light curves with background (host) emission 
subtracted (see Figure~\ref{wiselc}).

We consider the flux at certain epoch with signal to noise ratio (S/N) 
higher than 3 as a robust detection of variability. 
If none epoch satisfies the condition, we put a $3\sigma$ upper limit 
of the fluxes in which the $\sigma$ is determined by the mean errors of all epochs. 
According to this criterion, 11 TDEs show variability in either W1 or W2 band 
at one or multiple epochs.
Among them, we note that the IR echoes of ASASSN-14li at the first two epochs
has been reported by \citet{Jiang2016}. 
PTF-09ge and PTF-09axc only shows $3\sigma$ signal in W1 band but not in W2 band, 
that is also consistent with the results given by \citet{vV2016}.
All of the measurements are presented in Table~\ref{irfit}.

\begin{figure*}[b]
\centering
\subfigure{
\includegraphics[width=16cm]{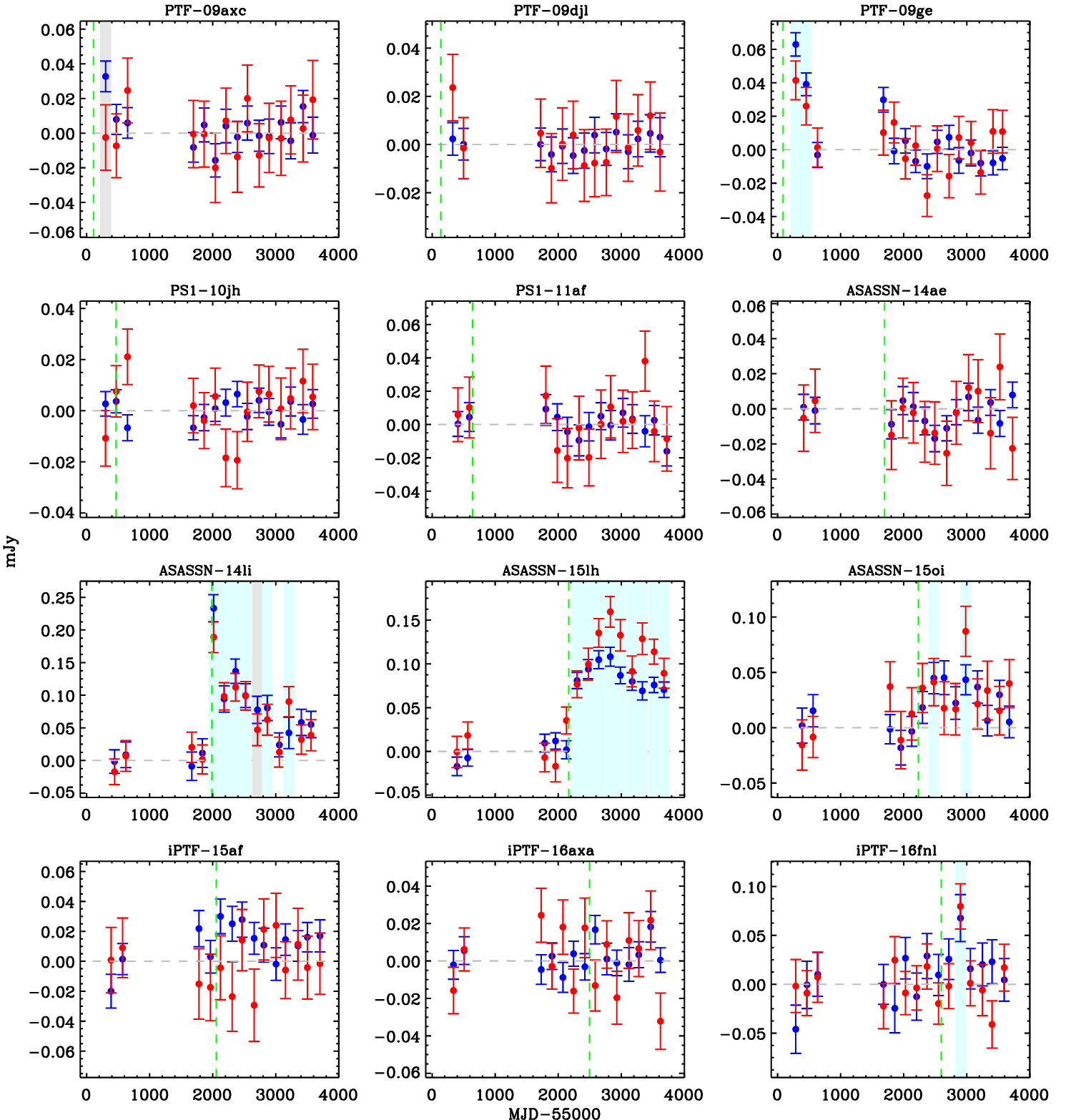}}
\caption{
The W1 (3.4$\mu$m, blue) and W2 (4.6$\mu$m, red) light curves of TDEs.
The fluxes (in unit of mJy) are measured with PSF photometry on the 
difference images. The green dashed lines mark the time of the optical peak.
We have highlighted the epochs with robust IR echo detections 
(dust emission measurements) in cyan shadow regions,
and the epochs with only $3\sigma$ W1 detection in grey shadow regions.
}
\end{figure*}
\addtocounter{figure}{-1}
\begin{figure*}[t]
 \centering
     \subfigure{
     \includegraphics[width=16cm]{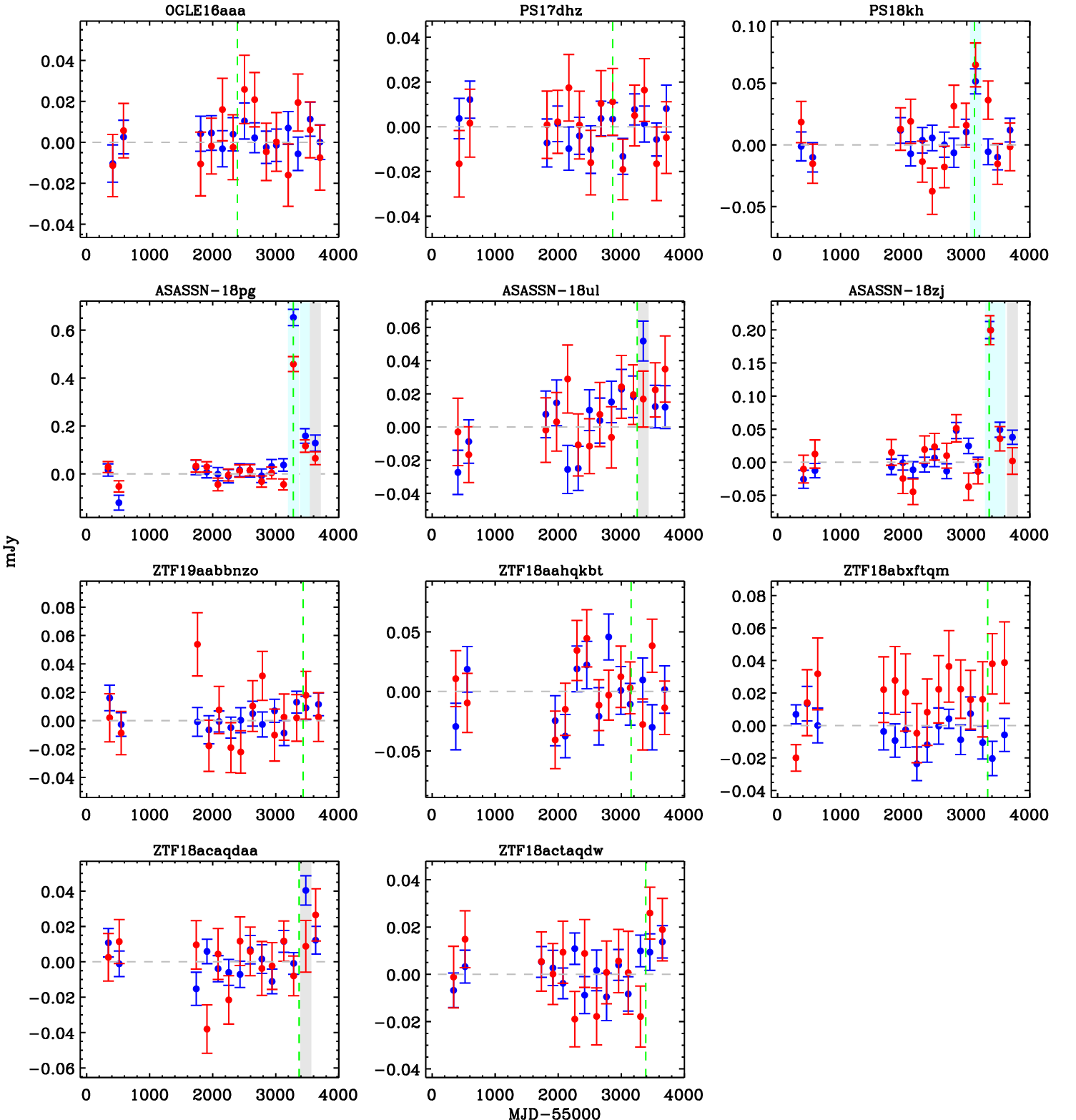}}
     \caption{continued}
	\label{wiselc}
\end{figure*}

\begin{deluxetable*}{rlrcccccc}
\tabletypesize{\scriptsize}
\setlength{\tabcolsep}{0.08in}
\tablecaption{\it IR emission of TDEs
\label{irfit}}
\tablewidth{0pt}
\tablehead{
\colhead{ID} & \colhead{Name} & \colhead{Days} & \colhead{$f_{W1}$} & \colhead{$f_{\rm W2}$} & \colhead{log$L_{\rm W1}$} & \colhead{log$L_{\rm W2}$} & \colhead{log$L_{\rm dust}$} & \colhead{$T_{\rm dust}$} \\
 & & & mJy & mJy & erg/s & erg/s & erg/s & K \\
\colhead{(1)} & \colhead{(2)} & \colhead{(3)}  & \colhead{(4)}  & \colhead{(5)}  & \colhead{(6)}  & \colhead{(7)}  & \colhead{(8)} & \colhead{(9)}}
\startdata
   1 &     PTF-09axc  &   182 & 0.033$\pm$0.009 & <0.054   & 42.00$\pm$0.12 & <42.10 & ... & ... \\
\hline
   2 &     PTF-09djl  & ...  & <0.022  & <0.043  & <42.27  & <41.95 & ... & ... \\
\hline
   3 &      PTF-09ge  &   200 & 0.063$\pm$0.007   & 0.041$\pm$0.012   & 41.74$\pm$0.05   & 41.42$\pm$0.12   & 42.22$\pm$0.68   &   1976$\pm$ 1235 \\
   3 &      PTF-09ge  &   362 & 0.039$\pm$0.007   & 0.026$\pm$0.011   & 41.54$\pm$0.08   & 41.22$\pm$0.19   & 42.00$\pm$1.03   &   1933$\pm$ 1806 \\
\hline
   4 &      PS1-10jh  & ...  & <0.015  & <0.034  & <42.04  & <41.77 & ... & ... \\
\hline
   5 &      PS1-11af  & ...  & <0.026  & <0.055  & <43.13  & <42.84 & ... & ... \\
\hline
   6 &   ASASSN-14ae  & ...  & <0.023  & <0.055  & <40.96  & <40.72 & ... & ... \\
\hline
   7 &   ASASSN-14li  &    29 & 0.233$\pm$0.021   & 0.189$\pm$0.024   & 41.30$\pm$0.04   & 41.07$\pm$0.05   & 41.48$\pm$0.28   &   1338$\pm$  276 \\
   7 &   ASASSN-14li  &   199 & 0.094$\pm$0.020   & 0.098$\pm$0.021   & 40.91$\pm$0.09   & 40.79$\pm$0.09   & 41.01$\pm$0.47   &   1019$\pm$  290 \\
   7 &   ASASSN-14li  &   387 & 0.137$\pm$0.019   & 0.112$\pm$0.021   & 41.07$\pm$0.06   & 40.85$\pm$0.08   & 41.27$\pm$0.42   &   1337$\pm$  407 \\
   7 &   ASASSN-14li  &   549 & 0.099$\pm$0.018   & 0.099$\pm$0.022   & 40.93$\pm$0.08   & 40.79$\pm$0.10   & 41.05$\pm$0.46   &   1073$\pm$  306 \\
   7 &   ASASSN-14li  &   743 & 0.078$\pm$0.021   & 0.047$\pm$0.024   & 40.82$\pm$0.12   & 40.47$\pm$0.22 & ... & ... \\
   7 &   ASASSN-14li  &   903 & 0.080$\pm$0.019   & 0.063$\pm$0.024   & 40.84$\pm$0.10   & 40.59$\pm$0.16   & 41.08$\pm$0.83   &   1440$\pm$  922 \\
   7 &   ASASSN-14li  &  1257 & 0.042$\pm$0.024   & 0.090$\pm$0.023   & 40.56$\pm$0.25   & 40.75$\pm$0.11   & 40.87$\pm$0.80   &    623$\pm$  208 \\
\hline
   8 &   ASASSN-15lh  &   112 & 0.081$\pm$0.009   & 0.077$\pm$0.015   & 43.07$\pm$0.05   & 42.90$\pm$0.09   & 43.36$\pm$0.39   &   1367$\pm$  332 \\
   8 &   ASASSN-15lh  &   265 & 0.094$\pm$0.011   & 0.100$\pm$0.018   & 43.13$\pm$0.05   & 43.02$\pm$0.08   & 43.41$\pm$0.36   &   1224$\pm$  247 \\
   8 &   ASASSN-15lh  &   405 & 0.105$\pm$0.010   & 0.135$\pm$0.016   & 43.18$\pm$0.04   & 43.15$\pm$0.05   & 43.46$\pm$0.24   &   1037$\pm$  123 \\
   8 &   ASASSN-15lh  &   562 & 0.108$\pm$0.011   & 0.160$\pm$0.018   & 43.19$\pm$0.04   & 43.22$\pm$0.05   & 43.51$\pm$0.22   &    943$\pm$   96 \\
   8 &   ASASSN-15lh  &   697 & 0.087$\pm$0.010   & 0.133$\pm$0.018   & 43.10$\pm$0.05   & 43.14$\pm$0.06   & 43.42$\pm$0.26   &    920$\pm$  106 \\
   8 &   ASASSN-15lh  &   857 & 0.080$\pm$0.010   & 0.091$\pm$0.018   & 43.06$\pm$0.05   & 42.98$\pm$0.08   & 43.33$\pm$0.37   &   1143$\pm$  224 \\
   8 &   ASASSN-15lh  &   993 & 0.069$\pm$0.010   & 0.129$\pm$0.018   & 43.00$\pm$0.07   & 43.13$\pm$0.06   & 43.40$\pm$0.29   &    814$\pm$   98 \\
   8 &   ASASSN-15lh  &  1153 & 0.076$\pm$0.009   & 0.114$\pm$0.014   & 43.04$\pm$0.05   & 43.08$\pm$0.05   & 43.36$\pm$0.25   &    931$\pm$  106 \\
   8 &   ASASSN-15lh  &  1286 & 0.070$\pm$0.009   & 0.089$\pm$0.017   & 43.01$\pm$0.05   & 42.97$\pm$0.08   & 43.29$\pm$0.36   &   1052$\pm$  186 \\
\hline
   9 &   ASASSN-15oi  &   245 & 0.045$\pm$0.014   & 0.041$\pm$0.021   & 41.34$\pm$0.14   & 41.17$\pm$0.23   & 41.28$\pm$1.31   &    985$\pm$  707 \\
   9 &   ASASSN-15oi  &   752 & 0.043$\pm$0.014   & 0.087$\pm$0.023   & 41.33$\pm$0.14   & 41.50$\pm$0.11   & 41.63$\pm$0.58   &    654$\pm$  148 \\
\hline
  10 &     iPTF-15af  & ...  & <0.033  & <0.065  & <41.65  & <41.33 & ... & ... \\
\hline
  11 &    iPTF-16axa  & ...  & <0.023  & <0.041  & <41.78  & <41.43 & ... & ... \\
\hline
  12 &    iPTF-16fnl  &   316 & 0.068$\pm$0.024   & 0.080$\pm$0.023   & 40.56$\pm$0.15   & 40.49$\pm$0.13   & 40.66$\pm$0.68   &    917$\pm$  349 \\
\hline
  13 &     OGLE16aaa  & ...  & <0.025  & <0.044  & <42.22  & <41.86 & ... & ... \\
\hline
  14 &       PS17dhz  & ...  & <0.026  & <0.044  & <41.85  & <41.47 & ... & ... \\
\hline
  15 &        PS18kh  &    17 & 0.052$\pm$0.010   & 0.065$\pm$0.018   & 41.75$\pm$0.09   & 41.71$\pm$0.12   & 41.82$\pm$0.60   &    814$\pm$  223 \\
\hline
  16 &   ASASSN-18pg  &     0 & 0.653$\pm$0.034   & 0.458$\pm$0.031   & 41.63$\pm$0.02   & 41.34$\pm$0.03   & 41.73$\pm$0.21   &   1461$\pm$  246 \\
  16 &   ASASSN-18pg  &   197 & 0.159$\pm$0.030   & 0.116$\pm$0.026   & 41.01$\pm$0.08   & 40.74$\pm$0.10   & 41.30$\pm$0.56   &   1554$\pm$  738 \\
  16 &   ASASSN-18pg  &   358 & 0.129$\pm$0.033   & 0.065$\pm$0.027   & 40.92$\pm$0.11   & 40.49$\pm$0.18 & ... & ... \\
\hline
  17 &   ASASSN-18ul  &    95 & 0.052$\pm$0.012   & 0.017$\pm$0.017   & 41.59$\pm$0.10   & 40.96$\pm$0.43 & ... & ... \\
\hline
  18 &   ASASSN-18zj  &    21 & 0.200$\pm$0.013   & 0.200$\pm$0.022   & 41.94$\pm$0.03   & 41.80$\pm$0.05   & 42.01$\pm$0.23   &   1033$\pm$  133 \\
  18 &   ASASSN-18zj  &   170 & 0.049$\pm$0.011   & 0.036$\pm$0.018   & 41.33$\pm$0.10   & 41.05$\pm$0.22   & 41.60$\pm$1.20   &   1582$\pm$ 1509 \\
  18 &   ASASSN-18zj  &   369 & 0.038$\pm$0.011   & 0.002$\pm$0.020   & 41.22$\pm$0.12   & 39.79$\pm$4.55 & ... & ... \\
\hline
  19 &  ZTF19aabbnzo  & ...  & <0.026  & <0.052  & <41.67  & <41.37 & ... & ... \\
\hline
  20 &  ZTF18aahqkbt  & ...  & <0.059  & <0.069  & <41.51  & <40.96 & ... & ... \\
\hline
  21 &  ZTF18abxftqm  & ...  & <0.030  & <0.060  & <41.71  & <41.40 & ... & ... \\
\hline
  22 &  ZTF18acaqdaa  &    91 & 0.040$\pm$0.008   & 0.009$\pm$0.015   & 42.67$\pm$0.09   & 41.87$\pm$0.72 & ... & ... \\
\hline
  23 &  ZTF18actaqdw  & ...  & <0.022  & <0.039  & <42.00  & <41.63 & ... & ... 
\enddata
\tablecomments{
Column~(1): Object ID in this paper.
Column~(2): Discovery Name of the TDE.
Column~(3): Rest-frame days since the optical peak.
Column~(4): W1 flux in unit of mJy.
Column~(5): W2 flux in unit of mJy.
Column~(6): W1 luminosity.
Column~(7): W2 luminosity.
Column~(8)-(9): Fitted dust luminosity and temperature. 
The contribution from the UV-optical components has been 
subtracted off during the fitting.}
Only the epochs with $3\sigma$ detection have been presented, while $3\sigma$ 
upper limits are given for TDEs showing $<3\sigma$ detections at all epochs.
We have not given the values of Column~(8) and (9) for epochs 
with unreliable measurements ($\tdust>2000~K$).
\end{deluxetable*}

\begin{figure}
\centering
\includegraphics[width=8.5cm]{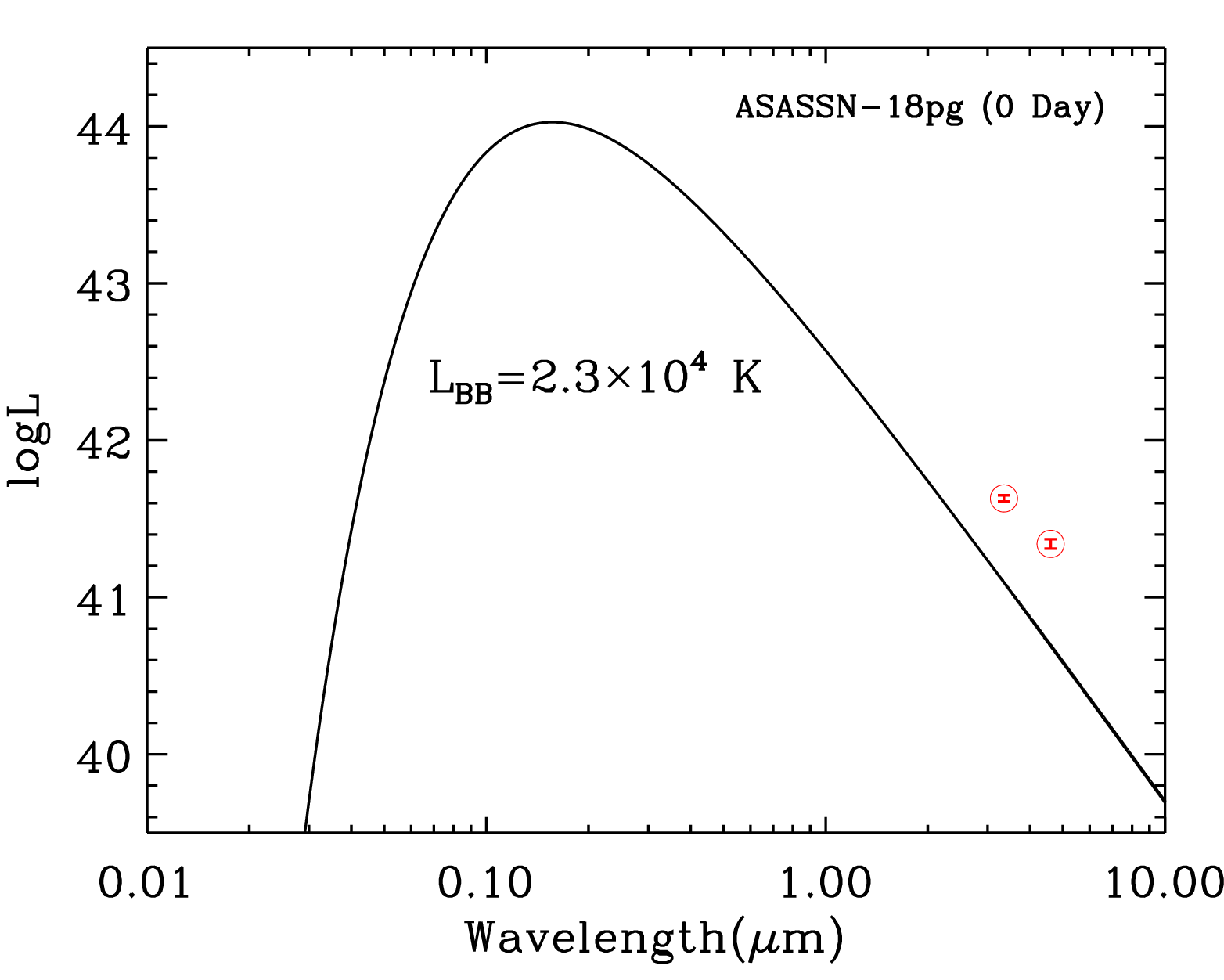}
\caption{
The SED of ASASSN-18pg at +0 days.
The black line is the blackbody spectrum determined from the optical-UV photometry
while the red circles denotes the observed luminosity at W1 and W2 band.
The observed IR emission shows evident excess relative to the UV-optical component.
}
\label{18pg}
\end{figure}

\begin{figure}
\centering
\includegraphics[width=8.5cm]{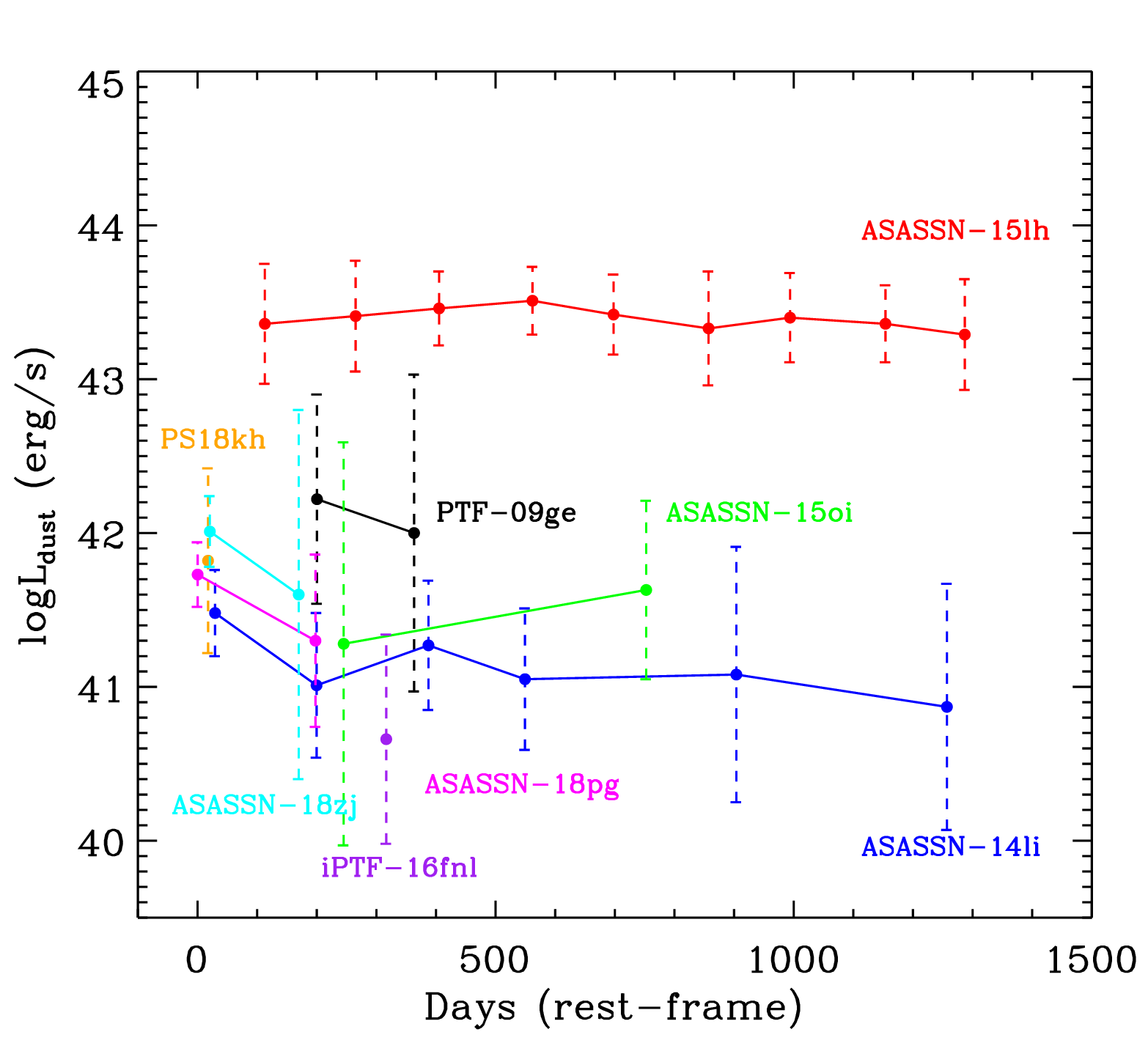}
\caption{
The fitted dust luminosity of the 8 TDEs with significant detections of IR echoes.
We show their evolution with time (the rest-frame days 
since the optical peak).
}
\label{lumlc}
\end{figure}

\begin{figure}
\centering
\includegraphics[width=8.5cm]{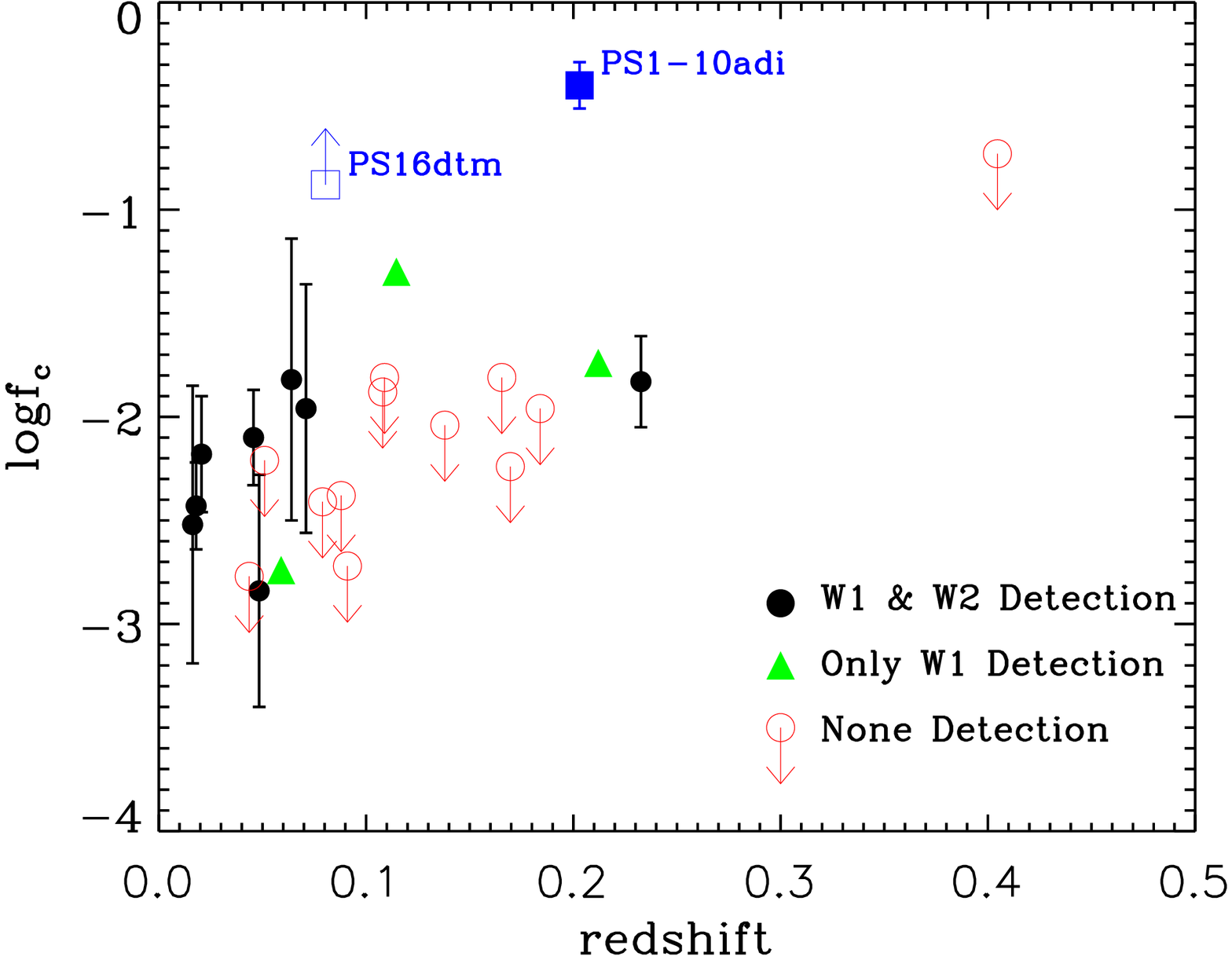}
\caption{
The logarithmic dust covering factor (log$f_c$) as function of redshift.
The black solid circles represent the 8 TDEs with reliable detection
in both W1 and W2 bands while the three with only detection in W1 band
are shown as green triangles.
The non-detected 12 sources are plotted with open red circles (upper limits).
As a comparison, we have also overplotted the two well sampled TDE candidates 
in AGNs, that is PS16dtm (\citealt{Blanchard2017}; \citealt{Jiang2017}) 
and PS1-10adi (\citealt{Kankare2017}; \citealt{Jiang2019}) with blue squares. 
The $f_c$ of PS16dtm is a lower limit since its MIR light curves is still rising.
Both $f_c$ are apparently higher than other TDEs in normal galaxies,
which is consistent with the AGN torus.
}
\label{cfz}
\end{figure}

\begin{deluxetable}{rlccccc}
\tabletypesize{\scriptsize}
\setlength{\tabcolsep}{0.04in}
\tablecaption{\it Dust Scale and Covering Factor}
\label{tbl-cf}
\tablewidth{0pt}
\tablehead{
\colhead{ID} & \colhead{Name} & \colhead{$t_{\rm peak}$} & \colhead{$R_d$} & \colhead{log$L_{\rm BB,peak}$} & \colhead{log$L_{\rm dust, peak}$} & \colhead{log$f_c$}  \\
 & & Days & pc & erg/s & erg/s & \\
\colhead{(1)} & \colhead{(2)} & \colhead{(3)}  & \colhead{(4)}  & \colhead{(5)} & \colhead{(6)} & colhead{(7)}
}
\startdata
   3 &     PTF-09ge  &  200 & 0-0.31    & 44.04$\pm$0.01 & 42.22$\pm$0.68 & -1.82$\pm$0.68 \\
   7 &  ASASSN-14li  &   29 & 0-0.17    & 43.66$\pm$0.02 & 41.48$\pm$0.28 & -2.18$\pm$0.28 \\
   8 &  ASASSN-15lh  &  562 & 0.35-0.60 & 45.34$\pm$0.04 & 43.51$\pm$0.22 & -1.83$\pm$0.22 \\
   9 &  ASASSN-15oi  &  752 & 0.49-0.78 & 44.47$\pm$0.04 & 41.63$\pm$0.56 & -2.84$\pm$0.56 \\
  12 &   iPTF-16fnl  &  316 & 0.12-0.41 & 43.18$\pm$0.03 & 40.66$\pm$0.67 & -2.52$\pm$0.67 \\
  15 &       PS18kh  &   17 & 0-0.16    & 43.78$\pm$0.02 & 41.82$\pm$0.53 & -1.96$\pm$0.60 \\
  16 &  ASASSN-18pg  &    0 & 0-0.15    & 44.16$\pm$0.01 & 41.73$\pm$0.17 & -2.43$\pm$0.21 \\
  18 &  ASASSN-18zj  &   21 & 0-0.17    & 44.11$\pm$0.01 & 42.01$\pm$0.21 & -2.10$\pm$0.23 \\
\hline
   1 &     PTF-09axc &  182 & 0.02-0.29 & 43.46$\pm$0.03 & 42.16$\pm$0.12 & -1.30$\pm$0.12* \\
  17 &   ASASSN-18ul &   95 & 0-0.22    & 44.48$\pm$0.04 & 41.74$\pm$0.10 & -2.74$\pm$0.11* \\
  22 &  ZTF18acaqdaa &   91 & 0-0.20    & 44.62$\pm$0.04 & 42.88$\pm$0.09 & -1.74$\pm$0.10* \\
\hline
   2 &     PTF-09djl & ...  & ...       & 44.42$\pm$0.04 & <42.46 & <-1.96 \\
   4 &      PS1-10jh & ...  & ...       & 44.47$\pm$0.07 & <42.23 & <-2.24 \\
   5 &      PS1-11af & ...  & ...       & 44.16$\pm$0.03 & <43.43 & <-0.73 \\
   6 &   ASASSN-14ae & ...  & ...       & 43.87$\pm$0.02 & <41.10 & <-2.77 \\
  10 &     iPTF-15af & ...  & ...       & 44.22$\pm$0.01 & <41.81 & <-2.41 \\
  11 &    iPTF-16axa & ...  & ...       & 43.82$\pm$0.02 & <41.94 & <-1.88 \\
  13 &     OGLE16aaa & ...  & ...       & 44.22$\pm$0.01 & <42.41 & <-1.81 \\
  14 &       PS17dhz & ...  & ...       & 43.82$\pm$0.05 & <42.01 & <-1.81 \\
  19 &  ZTF19aabbnzo & ...  & ...       & 44.56$\pm$0.06 & <41.84 & <-2.72 \\
  20 &  ZTF18aahqkbt & ...  & ...       & 43.87$\pm$0.01 & <41.66 & <-2.21 \\
  21 &  ZTF18abxftqm & ...  & ...       & 44.25$\pm$0.04 & <41.87 & <-2.38 \\
  23 &  ZTF18actaqdw & ...  & ...       & 44.21$\pm$0.29 & <42.17 & <-2.04 
\enddata
\tablecomments{
Column~(1): Object ID in this paper.
Column~(2): Discovery Name of the TDE.
Column~(3): The rest-frame days of the MIR peak since the optical peak (or first detection).
Column~(4): The range of the characteristic dust scale estimated from the 
time lag between MIR and optical emission, in which the uncertainties are mainly caused by 
the poor cadence of MIR light curves.
Column~(5): The peak (or first detection) blackbody luminosity of the optical-UV emission.
Column~(6): The peak dust luminosity with the UV-optical contribution subtracted off.
Column~(7): The dust covering factor estimated as $f_c=L_{\rm dust, peak}/L_{\rm BB,peak}$.
The \ldust\ of the first 8 objects, whose variability detections are reliable 
in both W1 and W2 bands, are calculated directly from the dust emission fitting.
The 3 objects in the middle, show >3$\sigma$ detection only in W1 band,
thus we have naively assumed that \tdust=1000~K, that is comparable with
the average of the first 8 objects. 
We denote their $f_c$  with asterisks, emphasizing their large uncertainties
and likely underestimated errors.
Lastly, only $3\sigma$ upper limits have been put for the rest 12 objects 
because their S/N are too low to infer the dust properties.
}
\end{deluxetable}

\subsection{Dust Emission}

After the detection of IR variability, the dust emission can be then estimated.  
Although the extrapolated emission of the UV-optical blackbody should be 
generally weak in the MIR bands, 
it could be not negligible at the very early stage of TDEs.
For instance, the IR variability of ASASSN-18pg is detected at almost the same
time with the optical peak. The expected logarithmic luminosity at W1 and W2 band
of the optical blackbody ($\tbb=2.3\times10^4~K$) is 41.09 and 40.69, that is only
0.54 and 0.65 dex lower than observations, respectively (see Figure~\ref{18pg}).

In order to minimize the contamination of optical blackbody, we then try
to subtract its contribution from the observed IR flux.  
We adopted the parameters listed in \citet{vV2020} to characterize the 
optical luminosity:
\begin{align}
L_{\rm bb}(t)=L_{\rm bb,peak}\left(\frac{t-t_{\rm peak}+t_{0}}{t_0}\right)^p , 
\end{align}
in which the relevant parameters can be found in Table~\ref{sample}.
Here we assume that blackbody temperature (\tbb) keeps constant 
to be the average of the first 100 days. 
Since ASASSN-15lh displays a non-monotonic declination with an exotic 
rebrightening feature in the UV-optical light curves (\citealt{Leloudas2016}), 
we have estimated its extension to the IR band from the measured luminosity 
at corresponding \wise\ epochs directly.

After flux correction, the dust temperature (\tdust) can be inferred as below.
\begin{align}
f_\nu &=\frac{1}{4\pi d_L^2}\int_{a_{\rm min}}^{a_{\rm max}} N(a) 4\pi a^2 Q_\nu(a) \pi B_\nu(T)~da
\end{align}

Following our previous works (e.g., \citealt{Jiang2021}), we simply
assume that the dust grains follow a MRN size distribution
(\citealt{Mathis1977}; see also \citealt{Draine1984}) as
$N(a)\propto a^{-3.5}$ with $a_{\rm min}=0.01\mu$m, $a_{\rm max}=10\mu$m
and an average density of $\rho=2.7\rm g~cm^{-3}$ for silicate grains.

Since PTF-09axc shows negative flux in W2 band at the epoch of $3\sigma$ 
W1 detection, we have ignored it in the following analysis.
In addition, the calculated \tdust\ of some epochs are unreasonably higher 
(with large errors) than the allowed temperature ($<1500-2000~K$) 
suppressed by dust sublimation (\citealt{Barvainis1987}; \citealt{Mor2012}).
Their are mainly caused by low S/N detection in W2 bands.
For this reason, we have abandoned the epochs with measured $\tdust>2000~K$
since their dust properties at corresponding epochs are obviously unreliable.
Accompanied with \tdust, the dust luminosity (\ldust) is obtained for the 8 TDEs 
with robust detection in both W1 and W2 bands (see Table~\ref{irfit}).  
The change of \ldust\ with time is shown in Figure~\ref{lumlc}.
The peak luminosity is mostly at orders of $10^{41}-10^{42}$~\lum\
except for ASASSN-15lh ($3.2\times10^{43}$~\lum), whose nature is 
still under hotly debated (\citealt{Dong2016}; \citealt{Leloudas2016}).

The spatial scale of dust can be roughly inferred from the time lag 
between the peak dust emission ($L_{\rm dust,peak}$) and 
primary optical-UV emission ($L_{\rm BB,peak}$).  
Apparently, the estimation have large errors due to poor cadence (half
year) of MIR light curves. As a conservative treatment, we put an error 
of 180 days (observer-frame) to the MIR peak time.
On the other hand, there are 7 TDEs which are detected only at the post-peak stage 
and their first detection epochs have been adopted as alternatives of the peak time
(marked as asterisks in Table~\ref{sample}). 
The errors introduced in this step might be minor as the luminosity at the 
first detection ($\sim10^{44}$~\lum) is comparable with the peak luminosity 
of other TDEs, indicating that the time of first detection is very close to their real peak.
It is also confirmed by the model fitting of ASASSN-14ae, ASASSN-14li and ASASSN-15oi 
(\citealt{Mockler2019}). 
Consequently, the time lag errors are mainly dominated by the peak time of dust emission. 
The corresponding dust scale from lags are presented as ranges in Table~\ref{tbl-cf}.
We can conclude that the dust revealed by IR echoes are all located at sub-pc, mostly at
$\lesssim0.2~pc$, even if it is hard to measure precisely with available data.

Another parameter widely adopted to characterize dust content is 
the covering factor of dust ($f_c$). 
Here we try to estimate it by the ratio of peak dust emission and
optical emission, that is $f_c=L_{\rm dust,peak}/L_{\rm BB,peak}$.
The 8 objects with $L_{\rm dust,peak}$ measurements show 
$f_c\sim0.01$ (see Table~\ref{tbl-cf}).
It needs to be emphasized that $f_c$ of ASASSN-15lh is on the same level
albeit with much higher luminosity.
We yet noticed that its MIR luminosity displays a much slower decay 
with that of the latest epoch remains comparable with the peak after 2 years.
If we employ the energy ratio as an alternative $f_c$ estimate of ASASSN-15lh, 
the integrated IR energy ($1.6\times10^{51}$~erg)
as of the end of 2019 yields out $f_c\sim0.09$ provided a total optical energy
$\sim(1.7-1.9)\times10^{52}$~erg (\citealt{Godoy-Rivera2017}).
The high IR luminosity of ASASSN-15lh is not only unusual in our TDE sample,
but also much higher than other superluminous supernova (L.M.Sun et al. in preparation).
Any future explanations of its nature must also account 
for the distinctive IR light curve successfully.

Regarding the three with only reliable detection in  W1 band,
we have made a very crude assumption of the \tdust\ ($1000~K$)
and have got a very similar $f_c$.
At the end, there are 12 objects left which have not been detected 
in W1 nor in W2 band. 
We choose to just give the $3\sigma$ upper limits of \ldust\ derived from 
the W1 luminosity (upper limits) as well as a fixed $1000~K$ \tdust.
The limits of \ldust\ are mainly dependent on the redshift of TDEs,
with closer TDEs having lower upper limits (see Figure~\ref{cfz}).
Taking the nearest (among the non-detected) TDE ASASSN-14ae (z=0.0436) 
as an example, \ldust<1.3$\times10^{41}$~\lum, 
resulting in $f_c$<$1.7\times10^{-3}$. 
To sum up, optical TDEs show $f_c\sim0.01$ or even less, which can be as low
as $f_c\lesssim10^{-3}$.
Because of the low dust content, only nearest TDEs (z<0.1 except for ASASSN-15lh)
show detectable IR echoes with \wise\ images (see Figure~\ref{cfz}).

\section{Conclusion and Discussion}

The pc-scale environment, namely the gas and dust, around SMBHs plays an
important role in understanding the triggering mechanism of AGNs.
Nevertheless, there is hitherto no statistical comparison of the environments
between active and quiescent galaxies.
The efforts are mainly hindered by the unachievable resolving power.
By means of IR echoes of TDEs, we now have the great opportunity to 
take a snapshot of dust at sub-pc scale in normal galaxies
(\citealt{Lu2016,Jiang2016,vV2016}). 
The study presented here has reviewed IR echoes of all optical TDEs 
discovered in the past decade (2009-2018) 
taking full advantage of available data resources. 
Confident (>$3\sigma$) IR variability has been detected in 11 targets and 
dust emission has been measured in 8 among them.
The concerned sub-pc scale dust revealed by IR echoes shows 
covering factor $f_c\sim0.01$, with the caveat that other non-detected 
sources show likely even lower $f_c$. 
Our conclusion agrees nicely with the pilot study of two TDEs (\citealt{vV2016}) 
while with higher statistical significance owing to the increased sample size. 
It is noticeable that our sample has gone beyond $z>0.2$ for luminous TDE 
candidate (ASASSN-15lh) , any current generation of instruments have not resolved 
sub-pc scale dust at such high redshift to our knowledge. 
In one word, our work has further proved that IR echoes can serve as 
a novel and effective method to probe dust down to sub-pc scale around SMBHs,
which is especially unique for inactive galaxies and distant galaxies.

\subsection{SMBHs are Quiescent due to Lack of Gas Supplies}

The torus in AGNs can reprocess the UV-optical photons from accretion 
to IR band and its covering factor has been commonly
inferred from the the spectral energy distribution (SED) decomposition
of the primary and reprocessed emission.
Past works have shown that the torus covering factor is averagely close to
one half (e.g., \citealt{Fritz2006}; \citealt{Mor2009}; \citealt{Roseboom2013}).
Consequently, TDEs in AGNs must produce associated IR outburst 
as a result of dust echoes. This scenario is fully supported by the ubiquitously
detected luminous ($10^{43}$-$10^{44}$~\lum) IR echoes in AGN TDEs
(\citealt{Dou2017}; \citealt{Jiang2017,Jiang2019}; \citealt{Mattila2018}).
Furthermore, the $f_c$ derived from them is also generally consistent with the 
traditional SED fitting method (\citealt{Jiang2019}; see also Figure~\ref{cfz}).

The optical TDEs considered in this work all happened in inactive galaxies
(including LINERs). They all show weak or non-detected IR echoes
with $f_c\lesssim0.01$, that is more than one order of magnitude lower than AGNs,
corroborating the conclusion given in \citet{vV2016}.  
It implies that either the pc-scale dust is quite sparse or 
it is concentrated on a geometrically thin and flat disk.
In any case, such few dust is definitely not sufficient to form a standard torus.
In comparison, the $f_c$ of the circumnuclear ring in the 
Galactic center is 0.12 inferred from parameters given by \citet{Lau2013}.
We may thus conclude that the $pc$-scale dust of normal galaxies,
represented by the optical TDE hosts,
is not only much less than AGNs, but also less than the Milky Way.
The selection effect is minimal here as they are all optically discovered 
without any prior IR information. 
In some theoretical models, AGN radiation pressure is necessarily involved
to both produce the torus toroidal structure and maintain its thick structure
(e.g., \citealt{Krolik2007}; \citealt{Wada2012}).
One may wonder if torus remains there when an AGN is turned off.
Our results hint that the torus, which is first introduced into the 
AGN unification as a toy model, may only exist in AGNs but not normal galaxies.
Likewise, the dusty wind in the polar direction should be also absent when
the SMBH is inactive, otherwise the polar dust will also 
responds to the TDE as an notable IR echo (e.g., \citealt{Mattila2018}).

In other words, SMBHs are dormant probably because of a shortage of gas 
in the vicinity instead of any formidable force to prevent gas flow to the BH.
AGNs seemingly to be easily triggered as long as the ambient pc-scale gas 
is rich. However, it is not the whole story.  
Apart from the TDEs analyzed in this work, which are captured by 
optical transient surveys, there is another class of TDE candidates selected 
by transient coronal line emitters and dust IR echoes 
(\citealt{Wang2012, Wang2018}; \citealt{Yang2013}; \citealt{Dou2016}).
They imply that SMBHs can be quiescent even if they lurk in ISM-rich environments.
Albeit we are aware of a selection effect of this technique itself,
it is worthwhile to further explore if the circumambient gas 
runs into a stone wall of losing angular momentum or they are on the eve 
of AGN phase (turn-on AGNs, e.g., \citealt{Gezari2017}; \citealt{Yan2019})
in the future.

\subsection{Implications to Demography of TDE Hosts}

One of the most puzzling open questions in TDE field is that optical TDEs
show an unexpected preference in post-starburst (or so-called E+A) galaxies,
with the rate  elevated by a factor of $\sim100$ (e.g. \citealt{Arcavi2014}; 
\citealt{French2016, French2020b}).
Scenarios which may contribute to the rate enhancement,
such as SMBH binaries (\citealt{Chen2009}, \citealt{Coughlin2019}),
central stellar overdensity (\citealt{French2020a})
and radial velocity anisotropy (\citealt{Stone2018}) have been proposed out.
However, those scenarios can not well address why TDEs are absent in galaxies with 
occurrent intense star formation (\citealt{Guillochon2017NA}).

The serendipitous discoveries of obscured TDEs in ultra-luminous infrared galaxies 
(ULIRGs) by IR echoes indicate promisingly that the absence of TDEs 
in star-forming (SF) systems is at least partly due to dust attenuation (\citealt{Tadhunter2017}; \citealt{Mattila2018}; \citealt{Kool2020}; see also \citealt{Sun2020}).
Actually, the TDE event rate of ULIRGs is estimated to be even
higher than post-starburst galaxies, which could be as high as
$\sim10^{-2}~{\rm gal}^{-1}~{\rm yr}^{-1}$ 
(\citealt{Tadhunter2017}, \citealt{Kool2020}).
Aside from ULIRGs, ordinary SF galaxies lying at the main sequence 
can also contain plenty of dust in the galactic nucleus.
A recent modeling of optical TDE detections in surveys indeed suggest that 
the dust obscuration is crucial for suppressing the TDE detection rate in 
SF galaxies while the unusual preference for post-starburst hosts 
can not be entirely explained (\citealt{Roth2021}). 
We caution that their estimate of extinction based on Balmer decrement have 
two caveats. First, the dust distribution is usually not uniform but 
the extinction of TDEs is only dependent on the dust along the line of sight.
Second, we still lack the knowledge of dust extinction in the galactic nucleus, 
which is likely different from the SF regions.
Thus further study on the impact of dust extinction must take the caveats
into consideration.

\citet{Jiang2021} has performed a blind search 
of MIR outburst in nearby galaxies and has yielded out a considerable number 
of TDE candidates in SF galaxies. 
Their peak MIR luminosity ($10^{42}-10^{44}$~\lum) is
much higher than the dust echoes revealed in optical TDEs.
This work and other progresses of TDE search by means of dust echoes suggest 
that the IR band is an efficient wavelength to unveil TDEs 
embedded in dusty environment.
In contrast, the optical search is only prone to uncover TDEs in SMBHs
with very low dust covering factor ($\lesssim0.01$).
If that is true, it on the other hand indicates that E+A galaxies might be a
special type of galaxies with the lowest amount of dust in the nuclear region.
The dust content in other inactive galaxies, like Milky Way, could be
significantly higher than E+A galaxies.
In the near future, we have the opportunity to obtain a panoramic picture of the 
pc-scale environment of SMBHs with IR echoes of TDEs in various types of galaxies.
\\

\acknowledgements

We are grateful to the anonymous referee for his/her careful reading
and valuable comments, which have greatly improved the paper.
We thank Dr. Aaron Meisner for his nice help of using unWISE images.
This work is supported by the NSFC (12073025,11833007,11421303), 
Joint Research Fund in Astronomy (U1731104) under cooperative agreement
between the NSFC and the CAS.
This research makes use of data products from the
\emph{Wide-field Infrared Survey Explorer}, which is a joint
project of the University of California, Los Angeles,
and the Jet Propulsion Laboratory/California Institute
of Technology, funded by the National Aeronautics and
Space Administration. This research also makes use of
data products from {\emph{NEOWISE-R}}, which is a project of the
Jet Propulsion Laboratory/California Institute of Technology,
funded by the Planetary Science Division of the
National Aeronautics and Space Administration.

\end{document}